\documentstyle[12pt]{article}

\begin{document}
\setlength{\oddsidemargin}{20 mm}
\setlength{\topmargin}{-10 mm}
\setlength{\textwidth}{17.0cm}
\setlength{\textheight}{23.0cm}
\renewcommand {\baselinestretch}{1.5}
\baselineskip 18 pt plus 1pt minus 1pt
\setcounter{page}{0}
\begin{titlepage}
\vspace*{-2 cm}
\begin{flushright}CEBAF-TH-92-26\end{flushright}\vspace*{3 cm}
\begin{center}
ELECTROMAGNETIC MASS  SPLITTINGS IN \\
HEAVY MESONS\vspace{1.5cm}\\
 J. L.  Goity \vspace*{5mm}\\
 {\it  Continuous Electron Beam Accelerator Facility\\
        Newport News, VA 23606, USA.}\vspace*{2 cm}\\
\end{center}
\begin{abstract}
The electromagnetic contribution to the isomultiplet mass splittings
of heavy mesons  is reanalyzed within the framework of the
heavy mass expansion.
 It is shown that the leading term in the expansion is given to
a good approximation by the elastic term. $1/m_{Q}$-corrections
can only be estimated, the main source of uncertainty now being inelastic
contributions. The $1/m_{Q}$-corrections to the elastic term
turn out
to be relatively small in both D  and B pseudoscalar mesons.
\end{abstract}
 \vspace*{1.6cm}
\begin{center}
November 1992
\end{center}
\end{titlepage}
\newpage
\setcounter{page}{1}
\section{Introduction}
The measurement of mass splittings in heavy pseudoscalar and vector mesons
has attained
a substantial degree of precision in the D-system $\cite{PDG,CLEO}$ and
in the B-system is  expected to improve  over the presently available
data $\cite{Bmesondata}$. In particular, the isospin splitting in D and
${\rm D^{\ast}}$ mesons is  now known within an error of $\pm 0.2 $ MeV
$\cite{CLEO}$, in B mesons  the error is larger  and about
$\pm 1$ MeV $\cite{Bmesondata}$ and, in ${\rm B^{\ast}}$  mesons its
 experimental determination is an important task still
to be accomplished.  The theoretical understanding of
isospin splittings is less satisfactory. Neither the electromagnetic
nor the quark mass contributions can be determined to a similar
degree of precision. In both cases, the task boils down to a strong
interaction problem where the extraction of a result cannot be
at present achieved without assumptions and, to some extent, modeling
$\cite{Fritzsch,quarkmod,BoalWright,GasserLeutwylerPR,GoityHou,Rosner}$.
Surprisingly good agreement with data was found $\cite{GoityHou}$
by an analysis in the limit of infinite heavy quark mass using a linear
interpolation to determine the contribution  $\Delta^{m}M$
to the splitting caused by the
mass difference of the u and d quarks, and the elastic approximation to the
electromagnetic contribution $\Delta^{\gamma}M$
 with a VMD model for the elastic form
factors associated with the light quark components of the electromagnetic
current.
  There are, however, unanswered questions concerning each of the
assumptions made. In the present work, we  reconsider   $\Delta^{\gamma}M$
in the  light of the heavy mass expansion. We  consider models
beyond  VMD  and give an estimate of
 corrections of order $1/m_{Q}$. Some conclusions of general character
are drawn from the analysis.

\section{Electromagnetic mass  splittings: pseudoscalar mesons}

The QED contribution to the mass of a hadron H is given to order
$\alpha$ by:
\begin{equation}
\delta^{\gamma} M_{H}
=
\lim_{\Lambda \rightarrow \infty}
\frac{i\; \alpha}{2 \,M_{H} \;(2 \pi)^{3}}
\int d^{4} q \; {\rm D}_{\mu\nu}(q)\, {\rm T}_{H}^{\mu\nu} (q, P)
\, \left[\frac{\Lambda ^{2}}{\Lambda^{2}-q^{2}}\right]\,
+
\delta^{\gamma} M_{H}^{\rm c.t.}(\Lambda )
\end{equation}
where  $P^{2}=M_{H}^{2}$, $-i\,{\rm D}_{\mu\nu}(q)$ is the photon
 propagator in an arbitrary gauge,  the
tensor ${\rm T}_{H}^{\mu\nu} $  the covariant forward
amplitude for virtual Compton scattering off H,  given by
 \begin{equation}
{\rm T}^{\mu\nu}_{H} (q, P)\equiv i\,\int \; d^{4} x \;e^{i\,qx} \;
\langle H, P, \sigma\mid  T \;j^{\mu}(x) j^{\nu}(0) \mid H, P, \sigma \rangle
{}~~~~,
\end{equation}
where $\sigma$ denotes the polarization of the hadron, and
$\delta^{\gamma} M_{H}^{\rm c.t.}(\Lambda )$ is the counterterm needed to
render
$\delta^{\gamma} M_{H}$ U.V. finite.
 The customary covariant normalization of states is used.
In this work, we focus on heavy mesons containing  a light
quark (u, d) and a heavy antiquark
($\bar {\rm c}$, $\bar {\rm b}$). $H$  and $H^{\ast}$ will denote respectively
the
pseudoscalar and vector mesons.
The electromagnetic current pertaining to our analysis is
conveniently decomposed as follows:
\begin{eqnarray}
j^{\mu}
&=&
J_{3}^{\mu}+ \frac{1}{2}\, J_{Y}^{\mu}+ e_{Q}\,J_{Q}^{\mu} \nonumber \\
J_{3}^{\mu}
&=&
\frac{1}{2}\; (\bar u \gamma^{\mu} u -\bar d\gamma^{\mu} d)\\
J_{Y}^{\mu}
&=&
\frac{1}{3}\; (\bar u \gamma^{\mu} u +\bar d\gamma^{\mu} d)\nonumber \\
J_{Q}^{\mu}
&=&
\bar Q \gamma^{\mu} Q \nonumber
\end{eqnarray}
where $Q$ denotes the heavy quark with charge $e_{Q}$.

 Since QCD plus QED is a renormalizable  and
parity conserving theory, the counterterm $\delta^{\gamma} M_{H}^{\rm
c.t.}(\Lambda )$
can only involve the matrix elements of
even parity  gauge invariant operators of dimension four or less
$\cite{Collins}$.
At order $\alpha$, the only such operators are $G^{\mu\nu}G_{\mu\nu}$
($G^{\mu\nu}$ is the
QCD field strength),
$\bar q_{i} q_{i}$ and $\bar Q Q$. The quark bilinear operators must appear
multiplied by a coefficient proportional to the respective quark mass as
demanded
by a consistent chiral limit. In this limit, the difference of mass shifts
  relevant to this work
 \begin{equation}
\Delta^{\gamma}M_{H} \equiv \delta^{\gamma} M_{H_{d}}-\delta^{\gamma} M_{H_{u}}
\end{equation}
becomes U.V. finite. This results from the cancellation of the gluon-
and heavy quark-
operator contributions in the difference.
 Throughout this work, we take the chiral limit;
this introduces an error in $\Delta^{\gamma}M_{H} $ of order $\alpha
m_{q}/\Lambda_{QCD}$.
 The $\bar s \gamma^{\mu} s$ component of the electromagnetic current
drops off when considering (4). For this reason it was
not included in (3).

In this section we discuss the case of pseudoscalar mesons. Following
Cottingham $\cite{Cottingham}$, one can perform in (1) an Euclidean rotation
in the $\nu=\frac{P.q}{M_{H}}$ complex plane and express:
\begin{equation}
\Delta^{\gamma}M_{H} =  \frac{ \alpha}{2 \,M_{H} \;(2 \pi)^{2}}\;\int
 \frac{d Q^{2}}{Q^{2}}
\;\int_{-Q}^{Q} \;d\nu\;\sqrt{Q^{2}-\nu^{2}}\,\; {\mbox{\Large
$\tau$}}^{\mu}_{\mu} (Q^{2},-i\,\nu)
\end{equation}
where $Q^{2}=-q^{2}$, and:
\begin{equation}
{\mbox{\Large $\tau$}}^{\mu \nu}\equiv T_{H_{d}}^{\mu\nu}- T_{H_{u}}^{\mu\nu}
\end{equation}

It is convenient to separate ${\mbox{\Large$ \tau$}}^{\mu \nu}$ into
two pieces, one referring only to
the light quark clouds and the other to the light quark cloud and the heavy
quark:
\begin{eqnarray}
{\mbox{\Large $\tau$}}^{\mu \nu}
\!\!&=&\!\!
{\mbox{\Large $\tau$}}_{A}^{\mu \nu}+
{\mbox{\Large $\tau$}}_{B}^{\mu \nu} \\
{\mbox{\Large $\tau$}}_{i}^{\mu \nu}
\!\!&=&\!\!
i\;\int \,d^{4} x\; e^{i\,qx}
\left(\langle H_{d}\mid T(J_{i}^{\mu}(0) \, J_{3}^{\nu}(x)
+J_{3}^{\mu}(0) \, J_{i}^{\nu}(x))\mid H_{d}\rangle - d
\rightarrow u \right)
\nonumber
\end{eqnarray}
where $i=A,B$, and $J_{A}^{\nu}\equiv \frac{1}{2} J_{Y}^{\nu}$
and $J_{B}^{\nu}\equiv e_{Q}J_{Q}^{\nu}$.

The lack of experimental access to the absorptive part of
${\rm T}_{H}^{\mu\nu}$, which would permit the reconstruction of
${\mbox{\Large $\tau$}}^{\mu \nu}$, compels us to resort to some
approximations.
The main approximation will consist in taking only the elastic terms in the
absorptive part of ${\mbox{\Large $\tau$}}^{\mu \nu}$. In the framework of the
heavy quark expansion, we will show that inelastic terms are of leading order
[${\cal O}(1)$] for $i=A$ and of order $1/m_{Q}$ for i=B.

The elastic contributions are conveniently studied in the
large $m_{Q}$ limit,  where the pseudoscalar and vector mesons belong
to a multiplet under the heavy quark symmetry $\cite{IsgurWise}$.
It is therefore appropriate  to keep as elastic  terms in the absorptive part
of $ {\mbox{\Large $\tau$}}^{\mu}_{\mu} $  not only the strictly elastic terms,
but also those inelastic terms which involve a M1 transition to
$H^{\ast}$ in the intermediate state.
 We can then write the elastic absorptive part
of ${\rm T}_{H}^{\mu\nu}$:
\begin{eqnarray}
W_{H;\,i}^{el\;\mu\nu}(q,P)
&\equiv&
\frac{1}{2} \;\sum_{n=H, H^{\ast}}\;\int \frac{d^{3}p_{n}}{2\,p_{n}^{0}}
\;(2\pi)\;\delta^{4} (p_{n}-P-q) \\
&\times &
\left[ \langle H, P\mid J_{i}^{\mu} (0)\mid n, p_{n}
\rangle\langle n, p_{n}\mid
J_{3}^{\nu} (0)\mid H, P \rangle +\begin{array}{cc}
i\leftrightarrow 3 \\ \mu\leftrightarrow \nu \end{array} \right]\nonumber\\
W_{i}^{el\;\mu\nu}&\equiv&
W_{H_{d};\,i}^{el\;\mu\nu}-W_{H_{u};\,i}^{el\;\mu\nu}\nonumber
\end{eqnarray}
with the matrix elements
\begin{eqnarray}
\langle H_{\beta}, P^{\prime}\mid J_{i}^{\mu} (0)\mid H_{\alpha}, P\rangle
&=&
\delta_{\alpha \beta} \;(P+P^{\prime})^{\mu} \;F_{i}(Q^2)\nonumber\\
\langle H_{\beta}, P^{\prime}\mid J_{3}^{\mu} (0)\mid H_{\alpha}, P\rangle
&=&
\tau_{\alpha \beta}^{3}\;(P+P^{\prime})^{\mu}\;F_{3}(Q^2)\nonumber\\
\langle H_{\beta}^{\ast}, \epsilon, P^{\prime}\mid J_{i}^{\mu} (0)
\mid H_{\alpha}, P\rangle
&=&
i \delta_{\alpha \beta}\; \epsilon^{\mu\nu\rho\sigma}\;\epsilon_{\nu}
P_{\rho}P^{\prime}_{\sigma}\; g_i(Q^2) \\
\langle H_{\beta}^{\ast}, \epsilon, P^{\prime}\mid J_{3}^{\mu} (0)
\mid H_{\alpha}, P\rangle
&=&
i \tau_{\alpha \beta}^{3} \;\epsilon^{\mu\nu\rho\sigma}\;\epsilon_{\nu}
P_{\rho}P^{\prime}_{\sigma} \;g_3(Q^2)\nonumber\\
F_{Y}(0)=\frac{1}{3}
&
F_{3}(0)=\frac{1}{2}& ~~~~~~~ \alpha,\;\beta=u,\;d\nonumber
\end{eqnarray}
where $\epsilon_{\nu}$ is the polarization four-vector of $H^{\ast}$.
In the large $m_{Q}$ limit
one has $\cite{IsgurWise}$:
\begin{eqnarray}
F_{Q}(Q^2)
&=&
 -e_{Q}\;\xi(v.v^{\prime})\\
g_{Q}(Q^2)
&=&
e_{Q}\;\frac{\xi(v.v^{\prime})}{M_{H}} ~~~~~~~ \xi(1)=1\nonumber
\end{eqnarray}
 where $\xi$ is the universal I-W  form factor, and $v$ denotes the
four-velocity of the meson.
The sign convention for $g_{Q}$  given by this equation implies that
sgn$F_{Y}$=sgn$g_{Y}$.

In the customary tensorial decomposition of the hadronic tensor, (8) and (9)
give
the following expressions  for the elastic structure
functions:
\begin{eqnarray}
W_{1;i}^{el\;\mu\nu}(q,P)
&=&
- 4 \pi\, M_{H}^{2}\, \delta(2\, M_{H} \,\nu-Q^{2}) \,\theta((P+q)^{0})
\,g_{i}(Q^{2})\,g_{3}(Q^{2})\;(\nu^{2}+Q^{2})\nonumber\\
W_{2;i}^{el\;\mu\nu}(q,P)
&=&
-4 \pi \,M_{H}^{2} \,\delta(2\, M_{H} \,\nu-Q^{2}) \,\theta((P+q)^{0})\\
  &\times&
\left[ 4\, F_{i}\,(Q^{2})\,F_{3}(Q^{2})+
g_{i}(Q^{2})\,g_{3}(Q^{2})\;\frac{\nu^{2}+Q^{2}}{1+\frac{Q^{2}}{4M_{H}^{
2}}}\right]\nonumber
\end{eqnarray}
At fixed $Q^{2}\geq 0$ the dispersion  integral over the variable $\nu$  gives:
\begin{eqnarray}
T_{1;i}^{\rm el}(Q^{2}, \nu)
&=&
-2\;g_{i}(Q^2)\,g_{3}(Q^2)
\frac{Q^{4}(1+\frac{Q^{2}}{4M_{H}^{2}})}{\frac{Q^{4}}{4M_{H}^{2}}
-\nu^{2}-i\,\epsilon \,Q^{2}}\\
T_{2;i}^{\rm el}(Q^{2}, \nu)
&=&
-2\left( 4F_{i}(Q^{2})\,F_{3}(Q^{2})+ Q^{2}\,g_{i}(Q^{2})\,g_{3}(Q^{2})\right)
\frac{Q^{2}}{\frac{Q^{4}}{4M_{H}^{2}}-\nu^{2}-i\,\epsilon \,Q^{2}}\nonumber
\end{eqnarray}
Using that:
\begin{equation}
 {\mbox{\Large $\tau$}}^{\mu}_{\mu}(q,\nu)=-3 \,T_{1}(Q^{2}, \nu)+
T_{2}(Q^{2}, \nu) (1+\frac{\nu^{2}}{Q^{2}})
\end{equation}
replacing in (5), and performing the integration over $d\nu$ we finally obtain:
\begin{eqnarray}
\Delta^{\gamma\;{\rm el}}M_{H}
&=&
 \Delta^{\gamma\;{\rm el}}_{A}M_{H}+\Delta^{\gamma\;{\rm el}}_{B}M_{H}\nonumber
\\
\Delta^{\gamma\;{\rm el}}_{i}M_{H}
&=&
\frac{\alpha}{\pi M_{H}} \;\int \,dQ^{2} \left(F_{i}(Q^{2})\,
F_{3}(Q^{2}) \;\left[(1+\frac{Q^{2}}{4\,M_{H}^{2}} )\;
(1-\sqrt{1+4 \frac{M_{H}^{2}}{Q^{2}}})+\frac{1}{2}\right]\right.\nonumber\\
&-&
\left.
g_{i}(Q^{2})\,g_{3}(Q^{2})\,\frac{Q^{2}}{2 }
\;\left[(1+\frac{Q^{2}}{4\,M_{H}^{2}}) \;
(1-\sqrt{1+4 \frac{M_{H}^{2}}{Q^{2}}})-\frac{1}{4}\right]\right)
\end{eqnarray}
To leading order in the heavy quark expansion the result becomes physically
very simple:
\begin{eqnarray}
\left. \Delta^{\gamma\;{\rm el}}M_{H}\right|_{0}
&=&
-\,\frac{4 \alpha}{\pi} \;\int_{0}^{\infty}
\;dQ \; \left(\frac{1}{2} \,F_{Y}(Q^{2})\,F_{3}(Q^{2})\right.\nonumber\\
&-&
\left.e_{Q}\; F_{3}(Q^{2})- \frac{Q^{2}}{4}
 \,g_{Y}(Q^{2})\,g_{3}(Q^{2})\right)
\end{eqnarray}
where the form factors are taken in the $m_{Q}\rightarrow\infty$ limit.
The first term is the difference between the Coulomb self-energies
of the light quark clouds with the u and d flavor quantum numbers,
the second term is the difference between the Coulomb interaction
of these clouds  with the heavy quark, and the third term
is the result of M1 transitions $H\leftrightarrow H^{\ast} \gamma $ mediated
by the light quark component of the e.m. current.
Since the relevant contributions
 to the integral in (15) stem from values of $Q^{2}$ of the order of
 light hadronic masses, one can neglect the
 $Q^{2}$-dependence  in  $F_{Q}(Q^{2})$
which amounts to an error of order $m_{Q}^{-2}$.
Since $g_{Q}$ is already of order $1/m_{Q}$, it does not appear in (15).

There are two types of $1/m_{Q}$-corrections, kinematic ones which are
obtained by expanding the terms in the square brackets in (14),
and dynamical ones which reside in the heavy mass dependence of the
form factors associated with the light quark currents. The latter are
hard to determine; charge conservation assures us that the
corrections vanish at $Q^{2}=0$ and  can only  affect the charge radius.
Kinematic
corrections due to the mass difference
$M_{H^{\ast}}-M_{H} \propto 1/m_{Q}$  are  of higher order.
One thus obtains the following $1/m_{Q}$-corrections:
\begin{eqnarray}
\left. \Delta^{\gamma\;{\rm el}}_{A}M_{H}\right|_{1}
&=&
\frac{3}{2} \frac{\alpha}{\pi \,M_{H}}\;
\int_{0}^{\Lambda}
\;dQ \left(Q\; F_{Y}(Q^{2})\,F_{3}(Q^{2})-\frac{1}{4}
Q^{3}\,g_{Y}(Q^{2})\,g_{3}(Q^{2})\right.
\nonumber \\
&-& \left. \frac{8}{3} F_{Y}(Q^{2}) \frac{\partial}{\partial\frac{ 1}{M_{H}}}
F_{3}(Q^{2})+ \frac{4}{3}\, Q^{2}\,g_{Y}(Q^{2})
\frac{\partial}{\partial \frac{1}{M_{H}}}g_{3}(Q^{2})\right)\\
\left. \Delta^{\gamma\;{\rm el}}_{B}M_{H}\right|_{1} &=&
-3\, e_{Q}  \frac{\alpha}{\pi \,M_{H}}\;
\int_{0}^{\Lambda}
\;dQ\; \left( Q\;F_{3}(Q^{2})-\frac{2}{3}
Q^{2}\,g_{3}(Q^{2})\right.
\nonumber\\
&-&
\left.\frac{4}{3} \,\frac{\partial}{\partial \frac{1}{M_{H}}}
F_{3}(Q^{2})\right)\nonumber
\end{eqnarray}
We have assumed here that the $m_{Q}$-dependence of $F_{Y}(Q^{2})$
 and $F_{3}(Q^{2})$ are the same. A similar assumption applies to the M1
form factors. In the limit where the light quark is taken as a constituent
quark, the term proportional to $g_{3}(Q^{2})$ in
$ \Delta^{\gamma\;{\rm el}}_{B}M_{H}$ is identified with the difference
between hyperfine
splitting. This
clarifies our inclusion of the M1 transition terms and also the sign
of the contribution.

It is important to notice  that the expansion in powers of $1/m_{Q}$
we have implemented corresponds to that of the HQET (heavy quark
effective theory). The  integration over $Q$ must then be cut off
at a scale $\Lambda << m_{Q}$, and the contributions which arise
from the domain
$Q>\Lambda $ must be included as local counter-terms of order $1/m_{Q}$.
These terms are, moreover, responsible for lifting the cut off
dependence of the results. A list of the possible counter-terms one can add has
been
given in ref $\cite{Rosner}$. The size of the coefficients
in front of these terms  is not known a priori. In the case of
the elastic contributions they could be determined once all the form factors
are known above the cut off.   Our ignorance about the counter-terms
is manifested, therefore, in the cut off dependence of our results,
which depends on the asymptotic
behavior of the intergrands in (16), and which turns out to be
more pronounced for $i=B$.

There are other terms of order $1/m_{Q}$ which have been disregarded; they are
the tadpole terms resulting from seagull type contributions to ${\rm
T_{H}^{\mu\nu}}$,
 and which must be present
to maintain gauge invariance. In particular, their strength is fixed at
$Q^{2}=0$ in terms of the electric
charge of the meson. This fact  has the particular consequence that at
order $1/m_{Q}$
the tadpole contributions to $\Delta^{\gamma}M_{H}$ and
$\Delta^{\gamma}M_{H^{\ast}}$
are equal.

 In conclusion, we are able to determine the $1/m_{Q}$-corrections of
kinematical origin, up to the mentioned cut off dependence, and those
originating
in hyperfine type interactions. Those residing in the form factors
will in any event be buried in our lack of knowledge of the form factors
themselves, and those of tadpole origin will be disregarded.

\section{Results}

For the sake of providing  a rough quantitative estimate of the
mass splitting  we now resort to a model for the form factors.
 In the case of the proton and the pion, the high $Q^{2}$ behaviour of the
form factors seems to pervade the  behavior at lower values
of $Q^{2}$, giving rise to the respective dipole and monopole
shapes. For heavy mesons the situation is  more involved.
Within the heavy quark expansion,  the asymptotic behavior
corresponds to $\Lambda_{QCD}^{2}\ll Q^{2}\ll m_{Q}^{2}$.
To leading order in $\alpha_{s}$, the  asymptotic behavior of the
charge  and M1 form factors coincide
 with those
of the relativistic hydrogen atom with $\alpha=\frac{3}{4} \alpha_{s}$:
\begin{eqnarray}
F(Q^{2}) \propto Q\,g(Q^{2})& \propto & Q^{- \delta(\alpha_{s})}\nonumber\\
\alpha_{s}&=&\alpha_{s}(Q^{2})
\end{eqnarray}
The exponent has the following values at some illustrative points:
$\delta(0)=3$, $\delta(0.3)=2.8$,
$\delta(0.5)=2.5$, $\delta(0.6)=2.2$, $\delta(0.64)=2$. For those values
of $\alpha_{s}$ where one can trust this estimate the form factors at
large $Q^{2}$
fall off faster than for the pion. Since the asymptotic behavior
is not definitely known, we choose the form factors to have
the following  form:
\begin{eqnarray}
3\, F_{Y}(Q^{2})=2 \,F_{3}(Q^{2})
&= &
\left(1+\frac {Q^{2}}{m_{\rho}^{2}}\right)^{-\kappa}\\
3\, g_{Y}(Q^{2})=2\, g_{3}(Q^{2})
&= &
\beta \;\left(1+\frac
{Q^{2}}{m_{\rho}^{2}}\right)^{-\kappa-\frac{1}{2}}\nonumber
\end{eqnarray}
The choice of $m_{\rho}$ as the relevant scale is
natural, and  likely to be  a reasonably good
 approximation. The constant $\beta$ determines the rate of the transition
$H^{\ast} \rightarrow H\gamma$. We will choose here a value suggested
by quark models $\cite{Rosner}$: $\beta \sim 3 \;{\rm GeV}^{-1}$. Here
$\kappa$ will be varied
between 1, which corresponds to the vector meson dominance model
$\cite{GoityHou}$, and 1.5, which asymptotically corresponds to
a loosely-bound light quark. Neglecting the $1/m_{Q}$ corrections to the
form factors  one obtains:
\begin{eqnarray}
 \Delta^{\gamma\;{\rm el}}_{A}M_{H}
&=&
-\frac{\alpha\,m_{\rho}}{2 \pi}
\left\{
\frac{\sqrt{\pi}}{3} \frac{\Gamma(2\kappa-\frac{1}{2})}{\Gamma(2\kappa)}\;
\left(1-\frac{\beta^{2}\,m_{\rho}^{2}}{8\kappa}\right)
\right.\nonumber\\
 &+&
\left.\frac{m_{\rho}}{32 \,M_{H}}\;\frac{\beta^{2}\,m_{\rho}^{2}-8\kappa}
{\kappa(2\kappa-1)}\right\}~~~~~~~~~~\kappa \geq \frac{1}{2}
\nonumber\\
\Delta^{\gamma\;{\rm el}}_{B}M_{H}
&=&
e_{Q}\;\frac{\alpha\,m_{\rho}}{\pi}
\left\{ \sqrt{\pi} \frac{\Gamma(\kappa-\frac{1}{2})}{\Gamma(\kappa)}\right.\\
&+&
 \left.\frac{3 m_{\rho}}{4 M_{H}}\;\left(
\frac{1}{1-\kappa}+\beta \,m_{\rho} \,\frac{\sqrt{\pi}}{3}
\;\frac{\Gamma(\kappa-1)}{\Gamma(\kappa+\frac{1}{2})}\right.\right.\nonumber\\
&+&
\left.\left.\left(\frac{m_{\rho}}{\Lambda}\right)^{2(\kappa-1)}\;
\frac{1}{\kappa-1}\;(1-\frac{2}{3}\beta \,m_{\rho})\right)\right\}\nonumber
\end{eqnarray}
For the chosen range of values for $\kappa$, only $\Delta^{\gamma\;{\rm
el}}_{B}M_{H}$
shows non-analytic contributions when setting \vspace{6 mm}
$\Lambda\sim M_{H}$.

 Numerical results for three different choices of
$\kappa$ are shown in the table. In all cases
$\Delta^{\gamma\;{\rm el}}_{A}M_{H}$ is only a fraction
of $\Delta^{\gamma\;{\rm el}}_{B}M_{H}$.
For $\beta=0$ (i.e.,
in the strictly elastic approximation), the  considered
$1/m_{Q}$-corrections to $\Delta^{\gamma\;{\rm el}}_{A}M_{H}$
 are small: 0.1 MeV for D mesons and 0.03 MeV for B mesons,
while the corrections to  $\Delta^{\gamma\;{\rm el}}_{B}M_{H}$
are larger:  -0.7 to  -0.5  MeV for D mesons and -0.3 to -0.1 MeV
for B mesons, and tend to suppress the splitting. For $\beta=3 $
GeV$^{-1}$, the additional contributions to
$\Delta^{\gamma\;{\rm el}}_{A}M_{H}$  and $\Delta^{\gamma\;{\rm
el}}_{B}M_{H}$ due to the M1 transitions
are of order $m_{Q}^{0}$ and $1/m_{Q}$ respectively. They lead to a suppression
of $\Delta^{\gamma\;{\rm el}}_{A}M_{H}$ by a factor 1/3 to 1/2; this,
however, affects the overall
mass splitting by less than 0.35 MeV. The corrections to
$\Delta^{\gamma\;{\rm el}}_{B}M_{H}$,
identified before to be of hyperfine nature, have the sign opposite to that
of the kinematical $1/m_{Q}$-corrections, and  they are about 0.8 to
0.5 MeV
for D mesons and about -0.3 to -0.2 MeV for B mesons.
The cut off dependence is a function of $\beta$:
$\left|\Delta^{\gamma\;{\rm el}}M_{H}\right|$ increases (decreases)
with increasing $\Lambda$ if
$\beta=0$ ($\beta=3 $ GeV$^{-1}$), and according with (19) the sensitivity
diminishes as $\kappa$ increases. We estimate that the cut off dependence
affects the elastic term of D and B mesons by $\pm 0.2$ Mev  and
 $\pm 0.1$ Mev  respectively. The dependence on $\beta$ is important
in D mesons, due to the fact that the dependences of
 $\Delta^{\gamma\;{\rm el}}_{A}M_{D}$ and $\Delta^{\gamma\;{\rm el}}_{B}M_{D}$
add up, while the contrary occurs in B mesons.
$\Delta^{\gamma\;{\rm el}}M_{D}$ increases by about 1 MeV as $\beta$ increases
from 0 to 3 GeV$^{-1}$, and
$\Delta^{\gamma\;{\rm el}}M_{B }$  varies by at most $\pm 0.1$ MeV in
this interval.
Overall, for the chosen value of
$\beta$, $1/m_{Q}$-corrections are small even for D mesons, and most of the
uncertainty in $\Delta^{\gamma\;{\rm el}}_{B}M_{D,B}$  resides in our
lack of  theoretical knowledge of the form factors.

At this point, it is appropriate to briefly address the  problem of
determining the total isospin splittings. $\Delta^{m}M_{H}$ is
determined from the mass difference $M_{H_{s}}-M_{H_{d}}$, which after
the inclusion of  chiral corrections
reads  $\cite{Goity}$:
\begin{equation}
M_{H_{s}}-M_{H_{d}}= C_{H} \;(m_{s}-m_{d})- \frac{3 \,g^{2}}{128\;F_{0}^{2}}
\;(2 \,M_{K}^{3} - M_{\eta}^{3})+{\cal O}(m_{s}^{2})
\end{equation}
where $g$ determines the $H^{\ast}\rightarrow H \pi$
amplitude, $F_{0}$=93 MeV is the pion decay constant in the chiral limit,
and $C_{H}$ is an unknown constant. Using the ratio
$\cite{GasserLeutwylerPR}$ $R\equiv (m_{s}-\hat m)/
(m_{d}-m_{u})\sim 43$, with ${\hat m}$ the average of the u and d quark masses,
one obtains:
\begin{equation}
\Delta^{m}M_{H}=\frac{1}{R-1/2}\;(M_{H_{s}}-M_{H_{d}}+
\frac{3 \,g^{2}}{128\;F_{0}^{2}}
\;(2 \,M_{K}^{3} - M_{\eta}^{3}) +{\cal O}(m_{s}^{2}))
\end{equation}
Thus, the leading chiral corrections, which are non-analytic in $m_{s}$,
tend to increase the  magnitude of $\Delta^{m}M_{H}$.
Clearly, (21) makes sense only as far as $g$ is small enough; the
corrections are large for $g>0.3$ (for determinations of  $g$
in the quark model and in QCD sum rules see refs. $\cite{Kogan}$ and for
 constraints  see refs. $\cite{Amundson}$). Otherwise, the unknown
${\cal O}(m_{s}^{2})$ terms
must also be included.   We conclude that there is a large degree of
theoretical
uncertainty in  $\Delta^{m}M_{H}$.

For $g=0$ and the ratio $R\sim 43$
one obtains $\Delta^{m}M_{D}\simeq 2.3$ MeV. The positivity of
the non-analytic term and the experimental result $\Delta M_{H}=4.8$ MeV
$\cite{CLEO}$ imply that $\Delta^{\gamma}M_{D}< 2.5$ MeV.

\section{Inelastic contributions}

The main  omission of the analysis presented here is that
of inelastic contributions. We only kept those involving
$H^{\ast}$ in the intermediate state. The inelastic contributions to
$\Delta^{\gamma}_{A}M_{H}$   appear at leading order in the heavy quark
expansion. Since $\Delta^{\gamma\;{\rm el}}_{A}M_{H}$ is substantially smaller
than $\Delta^{\gamma\;{\rm el}}_{B}M_{H}$, one might hope that they
will  lead to numerically small contributions to the mass shift.
As already mentioned, those considered amount to less than 0.35 MeV.
As a curiosity, in the case of the p-n electromagnetic
mass difference it has been noticed long ago
$\cite{GasserLeutwyler}$ that inelastic terms give only a modest
contribution. On the other hand, for $\Delta^{\gamma}_{B}M_{H}$ we can
 prove the following statement: {\it the inelastic contributions to
$\Delta^{\gamma}_{B}M_{H}$
are of order $1/m_{Q}$}. The proof is as follows:
inelastic terms in
the absorptive part of ${\mbox{\Large $\tau$}}^{\mu}_{\mu}$
involve products of matrix elements of the form:
\begin{equation}
\langle H,P\mid J_{3}^{\mu}(0) \mid P^{\prime}; X
\rangle \langle P^{\prime}; X \mid J_{Q}^{\nu}(0)\mid H,P \rangle
\end{equation}
where the intermediate state involves a heavy hadron with momentum
$ P^{\prime}$ and light hadrons denoted by $X$ carrying
momentum $P_{X}$. Let us analyze the matrix elements
separately. For $\langle P^{\prime}; X\mid J_{Q}^{\nu}(0)\mid H,P\rangle$
we use Bjorken's sum rule
$\cite{Bjorken}$:
\begin{equation}
1=\frac{1}{2}(1+v.v^{\prime}) \left| \xi(v.v^{\prime})\right|^{2}+
\int_{0}^{\infty} d\epsilon\;\omega_{\rm inel}(\epsilon, v.v^{\prime})
\end{equation}
where $\omega_{\rm inel}$ is given by:
\begin{eqnarray}
\omega_{\rm inel}(\epsilon, v.v^{\prime})
&=&
\sum_{n;X} ^{\;\;\;\;\;\;\;\prime}\int
\frac{d^{3}P^{\prime}}{(2\pi)^{3} \,2\, P^{\prime}_{0}}\;
(2\pi)^{4} \delta^{4}(P-q-P^{\prime}-P_{X}) \;\left|
\langle n, P^{\prime};X\mid J_{Q}^{\mu} \mid  H,P\rangle
\right|^{2} \nonumber\\ &+& Z-graph
\end{eqnarray}
where $n$ denotes a heavy hadron and the prime on the sum implies that
for $X=\mid 0\rangle$ one
must take $n\not=H$.
Since each term on the RHS is positive, and  the form factor satisfies the
normalization condition $\left| \xi(1)\right|=1$, taken
near the point of zero recoil, Eq. (23) implies  the Cabibbo-Radicati
type sum rule
 $\cite{Bjorken}$:
\begin{equation}
\int_{0}^{\infty} d\epsilon\;\omega_{{\rm inel}}(\epsilon, v.v^{\prime})=
\left(\frac{1}{2}+ 2\, \left|\xi\right|^{\prime}(1)\right)
(1-v.v^{\prime})
\end{equation}
which, supplemented with the assumption that $\xi^{\prime}$ is non-singular
at zero recoil and mild requirements of continuity for
$\omega_{\rm inel}$, implies that
\begin{equation}
\left|\langle n, P^{\prime}; X\mid J_{Q}^{\nu}(0)\mid H,P\rangle\right|
\propto (v.v^{\prime}-1)^{\eta}~~~~~~~~~~~\eta\geq \frac{1}{2}
\end{equation}

The behavior of $\langle H,P\mid J_{3}^{\mu}(0)\mid P^{\prime}; X \rangle $
for $v\not=v^{\prime}$ is determined as follows: in order to change the
velocity of the
heavy quark by a finite amount the large  space-like momentum transfer $q$
must flow as shown in the figure. The magnitude of the momentum is
$Q^{2}\sim 4 \,m_{Q}^{2}\,(v.v^{\prime}-1)$ in the cases where
$P_{X}^{2}\ll m_{Q}^{2}$,
which are indeed the case  in our problem.  As
illustrated in the figure, the  flow  of the momentum $q$  gives rise to a
suppression factor of the order of $1/Q^{3}$ stemming from the light quark and
gluon propagators. This clearly holds for  all possible diagrams contributing
to the matrix element. Thus:
\begin{equation}
\left|\langle H,P\mid J_{3}^{\mu}(0)\mid P^{\prime}; X \rangle \right|
\propto
\frac{1}{m_{Q}^{3}\,(v.v^{\prime}-1)^{3/2}}
\end{equation}
which is valid as far as $m_{Q}\,(v.v^{\prime}-1)^{1/2} >>
\Lambda_{QCD}$.  Thus,  according
to (26) and (27), in this regime
the product  (22) becomes proportional to $m_{Q}^{-3}\,(v.v^{\prime}-1)^{-1}$.
When
$m_{Q}\,(v.v^{\prime}-1)^{1/2}\sim \Lambda_{QCD}$ the LHS of (27)
is not suppressed, and  (22) becomes proportional to $m_{Q}^{-1}$ since
$(v.v^{\prime}-1)^{1/2} \propto m_{Q}^{-1}$ in (26). This latter case
corresponds precisely to the inelastic term, which we included as elastic,
with $H^{\ast}$ in the intermediate state (hyperfine term).
We therefore conclude that $ {\mbox{\Large $\tau$}}_{B;\,\mu\nu}^{\rm
inel}(q,\nu)$
is suppressed with respect to the strictly  elastic one by a factor
$m_{Q}^{-3}\,(v.v^{\prime}-1)^{-1}$, which upon replacement in (5) leads to the
stated $1/m_{Q}$ suppression for  $\Delta^{\gamma\;{\rm inel}}_{B}M_{H}$.

\section{Vector mesons}

Finally, a few  comments on vector mesons. Their isospin mass splittings
are identical to the ones of pseudoscalar mesons at leading order in $1/m_{Q}$.
They differ at order $1/m_{Q}$, where  a number of terms can be identified as
possible   contributions  $\cite{Rosner}$.
 In particular, for the electromagnetic
component there are the following  terms: the hyperfine term, which
breaks the heavy quark symmetry and contributes a factor (-1/3) times
the hyperfine term for the pseudoscalars (the relative factor is the same
as in quark models);
charge form factors $F_{3,Y}$ for pseudoscalar and vector mesons
differ at  this order (we expect this effect to be smaller than
the hyperfine term);  $1/m_{Q}$ corrections to the elastic dipole
form factor of  the  light quark components of the electromagnetic current
between vector meson states which lead to corrections to
$\Delta_{A}^{\gamma\,{\rm el}}
M_{H^{\ast}}$ (expected to be relatively small).
The  electric-quadrupole transitions give corrections of order $1/m_{Q}^{2}$,
and,
as mentioned before, the tadpole contributions are the same as for the
pseudoscalars
up to corrections of order $1/m_{Q}^{2}$.

  In D mesons  the data is
summarized by $\cite{PDG,CLEO}$:
\begin{eqnarray}
(a) \;\;\;\;\;\;\;\;\;\;\;\;\;\;\;\;\;\;\;\;\;\;\;\;\;\;\;\;
M_{D^{\ast}_{s}}-M_{D_{s}} &=& 141.5\pm 1.9\; {\rm MeV}\nonumber\\
M_{D^{\ast\,0}}-M_{D^{0}}
 &=& 142.12\pm 0.07\;
{\rm MeV}\nonumber\\
M_{D^{\ast\,+}}-M_{D^{+}}&=& 140.64\pm 0.1\; {\rm MeV}\\ & & \nonumber \\
(b) \;\;\;\;\;\;\;\;\;\;\;\;\;\;\;\;\;\;\;\;\;\;\;\;
M_{D^{\ast\,+}}-M_{D^{\ast\,0}}&=& 3.32\pm 0.1\; {\rm MeV}\nonumber\\
M_{D^{+}}-M_{D^{0}}&=& 4.8\pm0.1{\rm MeV}\nonumber
\end{eqnarray}
The results (a) show that the QCD hyperfine splitting is remarkably SU(3)
symmetric,
as emphasized in $\cite{GoityHou}$: there is a change of less than 2$\%$
as the quark masses vary from $m_{u,d}$ to $m_{s}$. Unless this is an accident
resulting from the fact that non-linear terms in $m_{s}$ are large,
one concludes that $\cite{GoityHou}$
$\Delta^{m}M_{D}=\Delta^{m}M_{D^{\ast}}$ to a good
degree of precision. Thus,
$\Delta M_{D^{\ast}}-\Delta M_{D}\simeq
\Delta^{\gamma}M_{D^{\ast}}-\Delta^{\gamma}M_{D}\simeq -1.48\pm0.15$ MeV,
which suggests a hyperfine term in $\Delta^{\gamma \,{\rm el}}M_{D}$
of about 1 MeV,  a value which is not very different from the ones
 we obtain, which for $\beta=3$ GeV$^{-1}$ vary between 0.5 to 0.8 MeV.
Clearly, the possibility of isolating the hyperfine contribution
provides  a useful constraint for further theoretical understanding of
$1/m_{Q}$ corrections.

\section{Conclusions}
Electromagnetic contributions to mass splittings in D and B mesons are
conveniently analyzed in the framework of the large mass expansion.
Leading order terms are determined, up to inelastic corrections to
$\Delta_{A}^{\gamma}M_{H}$, by elastic contributions which are
given in terms of the elastic form factors associated with the light
quark components of the electromagnetic current. A precise theoretical
determination
of the latter from QCD is thus very important. The $1/m_{Q}$
corrections suffer from uncertainties of different sorts, and a precise
treatment is
difficult. We have shown that the $1/m_{Q}$-corrections to elastic
contributions  can be estimated to a good extent, and, within the
 approximations used here, for pseudoscalar mesons they turn out to be
relatively small due to the fact
that kinematic and hyperfine corrections conspire to cancel each other.
For $\beta=3$ GeV$^{-1}$ the results are remarkably close to those
obtained at leading order. To fully pin down the corrections to
the elastic term knowledge of the $m_{Q}$ dependence of $F_{Y,3}$
and $g_{Y,3}$ is required, and moreover, the difficult-to-estimate
tadpole terms should be included. Much more work is needed
in order to  estimate the importance of the disregarded
inelastic terms.

Our main aim was to  conceptually clarify the analysis of electromagnetic
splittings, and therefore, we refrained from making definite
numerical claims, which are sensitive to the model chosen for the
elastic form factors and other assumptions. Our estimates for the
relative sizes of the
$1/m_{Q}$ corrections to  $\Delta^{\gamma\,{\rm el}}M_{H}$ should however
be taken  seriously.

\section{Acknowledgements}
We thank Simon Capstick, Michael Frank and Javier Gomez for useful and
informative
discussions, and especially Nathan Isgur for pointing out
a most important sign mistake and for comments on the manuscript.

\newpage

\begin{tabular}{|c|c|c|c|c|c|c|c|} \cline{1-8}
&  & \multicolumn {3}{c|}{ }&\multicolumn {3}{c|}
{}\\
&  & \multicolumn {3}{c|}{$\beta=0$ }&\multicolumn {3}{c|}
{$\beta=3 \;{\rm GeV}^{-1}$}\\
\cline{3-8}
{\large H}& $\kappa $ &  &  &  &  &  &  \\
& & $\Delta^{\gamma\;{\rm el}}_{A} M_{H}$ & $\Delta^{\gamma\;
{\rm el}}_{B} M_{H}$ & $\Delta ^{\gamma\;{\rm el}} M_{H}$ &
 $\Delta^{\gamma\;{\rm el}}_{A} M_{H}$ & $\Delta^{\gamma\;
{\rm el}}_{B} M_{H} $ & $\Delta^{\gamma\;{\rm el}} M_{H} $\\
&  & MeV & MeV & MeV & MeV & MeV & MeV
\\  \hline
& {\bf 1} & {\bf -0.39} & {\bf 3.0} & {\bf 2.6}  &  {\bf -0.12}  & {\bf 3.8}
&  {\bf 3.7$\pm$ 0.2}  \\
& & -0.47 & 3.7 & \fbox{3.2 }& -0.15 & 3.7 & 3.6 \\
\cline{2-8}
  & {\bf 1.25}  & {\bf -0.34}  & {\bf 2.3}  & {\bf 1.9}  & {\bf -0.15}  &
{\bf 2.9}  & {\bf 2.7$\pm$ 0.2} \\
{\large D} & & -0.40 & 2.8 & 2.5 & -0.18 & 2.8 & 2.6 \\
 \cline{2-8}
 & {\bf 1.5}  & {\bf -0.30}  & {\bf 1.9}  &
{\bf 1.6}  & {\bf -0.17}  & {\bf 2.4}  & {\bf 2.2$\pm$ 0.1} \\
& & -0.35 & 2.4 & 2.0 & -0.19 & 2.4 & 2.2 \\
\hline
& {\bf 1 } & {\bf -0.44}  & {\bf -1.6}  & {\bf -2.1}  & {\bf -0.14}  &
{\bf -1.9}  & {\bf -2.1$\pm$ 0.1}  \\
& & -0.47 & -1.9 &\fbox{ -2.4} &  -0.15 & -1.9 & -2.1 \\
\cline{2-8}
& {\bf 1.25}  & {\bf -0.38}  & {\bf -1.3}  & {\bf -1.7}  & {\bf -0.17}
& {\bf - 1.5}  & {\bf -1.6$\pm$ 0.1}  \\
{\large B} & & -0.40 & -1.4 & -1.8 & -0.18 & -1.4 & -1.6 \\
\cline{2-8}
&{\bf  1.5}  & {\bf -0.33}  &  {\bf -1.1}  & {\bf -1.4}  & {\bf -0.19}
& {\bf -1.2}  & {\bf -1.4$\pm $0.05} \\
& & -0.35 & -1.2 & -1.5 & -0.19 & -1.2 & -1.4 \\
\hline
\end{tabular}
\vspace*{6 mm}\\
\newpage

CAPTIONS\vspace*{10mm}\\

\parbox{4.5in}{{\bf Table:} {\it Results for electromagnetic mass splittings
of pseudoscalar mesons in the
elastic approximation as defined in the text. Bold type numbers refer to
$1/m_{Q}$ corrected values. The results correspond to the choice
$\Lambda\sim M_{H}$ in eq. (19), and the error shown in the last
column results from a change in $\Lambda$ by a factor of 2.
For comparison the results in the $m_{Q}
\rightarrow \infty$ limit are also shown. The numbers designated by
boxes  are the results of the VMD model $\cite{GoityHou}$.}}
\vspace*{15 mm}

\parbox{4.5in}{{\bf Figure:}  {\it The  diagram  shows the flow of the
large space like momentum $q$ in the process $H\rightarrow H'+X$  mediated by
the light quark component
of the electromagnetic current.}}


\newpage
\begin{thebibliography}{99}
\bibitem{PDG} Review of Particle Properties, Phys. Rev. D 45 (1992).
\bibitem{CLEO} S. Barlag et al., Phys. Lett. B 278 (1992) 480.\\
D. Bortoletto et al., Phys. Rev. Lett. 69 (1992) 2046.
\bibitem{Bmesondata} H. Albrecht et al., Z. Phys. C 48 (1990) 543.\\
J. Lee-Franzini et al., Phys. Rev. Lett. 65 (1990) 2947.\\
D.S. Akerib et al., Phys. Rev. Lett. 67 (1991) 1692.\\
D. Bortoletto et al., Phys. Rev. D 45 (1992) 21.
\bibitem{Fritzsch} H. Fritzsch, Phys. Lett. 63 B (1976) 419, and,
Phys. Lett. 71 B (1977) 429.
\bibitem{quarkmod}
A. De R\'ujula, H. Georgi and S.L. Glashow, Phys. Rev. D 12 (1975) 147.\\
S. Ono, Phys. Rev. Lett. 37 (1976) 655.\\
K. Lane and S. Weinberg, Rev. Lett. 37 (1976) 717.\\
W. Celmaster, Phys. Rev. Lett. 37 (1976) 1042.\\
L.H. Chan, Phys. Rev. D 15 (1977) 2478, and, Phys. Rev. Lett. 51 (1983) 253.\\
R.J. Johnson, Phys. Rev. D 17 (1978) 1459.\\
N. Isgur, Phys. Rev. D 21 (1980) 779.\\
S. Godfrey and N. Isgur, Phys. Rev. D 34 (1986) 899.
\bibitem{BoalWright} D.H. Boal and A.C.D. Wright, Phys. Rev. D 16 (1977) 1505.
\bibitem{GasserLeutwylerPR}J. Gasser and H. Leutwyler, Phys. Rep. 87 C (1982)
77.
\bibitem{GoityHou} J.L. Goity and W-S. Hou, Phys. Lett. B  282 (1992) 243.
\bibitem{Rosner} J.L. Rosner and M.B. Wise, {\it Meson Masses from SU(3)
and Heavy Quark Symmetry}, Report CALT-68-1807 (1992).
\bibitem{Collins} J.C. Collins, Nucl. Phys. B 149 (1979) 90.
\bibitem{Cottingham} W.N. Cottingham, Ann. of Phys. 25 (1963) 424.
\bibitem{IsgurWise} N. Isgur and  M.B. Wise, Phys. Lett. B 232 (1989) 113.
\bibitem{Goity} J.L. Goity, Phys. Rev. D 46 (1992) 3929.
\bibitem{GasserLeutwyler} J. Gasser and H. Leutwyler, Nucl. Phys. B 94 (1975)
269.
\bibitem{Bjorken} J.D. Bjorken, Lectures at the 18$^{th}$ Annual SLAC
Summer Institute in Particle Physics, SLAC-PUB-5389, (1990).\\
  J.D. Bjorken, I. Dunietz and J. Taron, Nucl. Phys. B 371 (1992) 111.
\bibitem{Kogan} N. Isgur and M.B. Wise, Phys. Rev. D 41 (1990) 151.\\
V.L. Eletsky and Ya.I. Kogan, Z. Phys. C 28 (1985) 155.
\bibitem{Amundson} J.A. Amundson et al., {\it Radiative ${\rm D}^{\ast}$-Decay
using Heavy Quark and Chiral Symmetry}, Report UCSD/PTH 92-31  (1992).\\
H.Y. Cheng et al., {\it Chiral
Lagrangians for Radiative Decays of Heavy Hadrons}, Report CLNS 92/1158
(1992).\\
P. Cho and H. Georgi, {\it Electromagnetic Interactions in Heavy Hadron
Chiral Theory},
report HUTP-92/A043 (1992).
\end{thebibliography}
\end{document}